\documentclass[runningheads]{llncs}
\usepackage{graphicx}
\graphicspath{{./images/}}
\DeclareGraphicsExtensions{.pdf,.jpeg,.png,.PNG,.JPEG,.eps,.pdf}
\usepackage{cite}
\usepackage{array}
\usepackage{microtype}
\usepackage{comment}
\usepackage{ctable}
\usepackage{amssymb}
\usepackage{listings}
\usepackage{subfig}
\usepackage{xfrac}
\usepackage{mathtools}
\usepackage{booktabs}
\usepackage{multirow}
\usepackage{url}
\usepackage{float}

\definecolor{gray97}{gray}{.97}
\definecolor{gray75}{gray}{.75}
\definecolor{gray45}{gray}{.45}

\makeatletter
\def\lst@PlaceNumber{\ifnum\value{lstnumber}=0\else
  \llap{\normalfont\lst@numberstyle{\thelstnumber}\kern\lst@numbersep}\fi}
\makeatother   
\lstdefinestyle{C} {
    language=C++,
    morekeywords={in,out,downto,IOCTL}
   }
\newcommand{\ignore}[2]{\hspace{0in}#2}

\begin{document}
\title{High-Performance Simultaneous Multiprocessing for Heterogeneous System-on-Chip}
\subtitle{Position Paper}
\titlerunning{High-Performance Multiprocessing with Heterogeneous System-on-Chips}
%
\author{Kris Nikov\inst{1}\and
Mohammad Hosseinabady\inst{1}\and
Rafael Asenjo\inst{2}\and
Andr\'es Rodr\'iguez\inst{2}\and
Angeles Navarro\inst{2}\and
Jose Nunez-Yanez\inst{1}
}
\authorrunning{K. Nikov et al.}
%
\institute{ University of Bristol, UK \\
              \email{\{kris.nikov, m.hosseinabady, j.l.nunez-yanez\}@bristol.ac.uk} \\
          \and
          	Universidad de M\'alaga, Spain \\
              \email{\{asenjo, andres, angeles\}@ac.uma.es}
}

\maketitle              
\vspace*{-7mm}
\begin{abstract}
This paper presents a methodology for simultaneous heterogeneous computing, named ENEAC, where a quad core ARM Cortex-A53 CPU works in tandem with a preprogrammed on-board FPGA accelerator. A heterogeneous scheduler distributes the tasks optimally among all the resources and all compute units run asynchronously, which allows for improved performance for irregular workloads. ENEAC achieves up to 17\% performance improvement \ignore{and 14\% energy usage reduction,} when using all platform resources compared to just using the FPGA accelerators and up to 865\% performance increase \ignore{and up to 89\% energy usage decrease} when using just the CPU. The workflow uses existing commercial tools and C/C++ as a single programming language for both accelerator design and CPU programming for improved productivity and ease of verification. 
\end{abstract}
\vspace*{-7mm}
\section{Introduction}

With the advent of Dark Silicon and the end of Dennard Scaling ~\cite{esmaeilzadeh2011dark}~\cite{shafique2014dark}, heterogeneous systems are seen as the way for the semiconductor industry to keep up with performance demands. This is not surprising since DSPs, GPUs and NPUs are already widely used coprocessors, however emerging fields such as cryptosecurity and artificial neural networks have also raised the demand for dedicated on-chip accelerators. With more of these being integrated in consumer devices, it is inevitable that eventually the trade-off of increase chip area will necessitate the reuse of silicon for task acceleration. FPGAs are set to play an important role in the future of Heterogeneous Computing for upcoming generations of SoCs having the ability to be reprogrammed with specific accelerators on-demand and within context switches. 

The methodology, called ENergy Efficient Adaptive Computing with heterogeneous architectures (ENEAC) presented in this paper aims to build upon existing tools and platform in order to develop a comprehensive solution to Heterogeneous Computing using CPUs and FPGAs. This paper presents a continuation of the work in Nunez-Yanez et al. \cite{nunez2019simultaneous}, and moves from the original Zynq embedded devices to add support for high-performance Zynq Ultrascale devices. The workflow is updated and evaluated on the Xilinx ZCU102 Development Platform with a larger FPGA device and 64-bit ARM processors. The entire methodology and usage tutorial are open-source and available online at~\cite{ENEAC}.

Key contributions include:
\begin{enumerate}
\item A framework for customising accelerator code and programming the FPGA using the Xilinx SDSoC development environment.
\item A scheduling algorithm which distributes the workload between the CPU cores and FPGA accelerators and ensures load balance among devices.
\item A custom platform that adds extensive interrupt management to enable the accelerators to work independently, which improves system throughput for irregular workloads.  
\end{enumerate}
\vspace*{-3mm}
\section{Related work}
Heterogeneous computing is not just a means to improve performance, but can also be highly effective in areas where minimizing energy usage is critical, such as embedded systems. The limitations of CPUs are highlighted by Chung et al.~\cite{micro10}, who demonstrate that over 90\% of the energy in a general-purpose processor is ``overhead''. There is a clear need to integrate more application-specific accelerators and current efforts to promote Heterogeneous Computing include the Heterogeneous System Architecture (HSA) Foundation~\cite{HSA}. They present a new integrated computational platform and associated software tools that allows distributed workload execution over a variety of processors from a single software source.

The majority of heterogeneous computing involves using a host processor, which controls the execution across the other compute units. ENEAC explores a horizontal collaborative approach, where the platform CPU also contributes to the task execution, thus improving performance compared to solely using the FPGA as an accelerator. The developed workflow includes commercially available tools from the Xilinx suite to enable quick adoption of new algorithms/workloads and easy system reconfiguration. A similar approach has been explored by Tsoi et al.~\cite{fpga10}. They focus on using multiple devices to showcase how the Nbody simulation can be successfully implemented on a heterogeneous system and use both FPGA and GPU to compute the same kernel on different portions of particles, which achieves a 22.7 times speedup compared to the CPU only version.

This work is a continuation of the methodology presented in ~\cite{nunez2019simultaneous}, however the workflow has been further optimised for irregular workloads and has been validated on a more complex platform. The key updates include the ability to schedule to a set of accelerators, programmed on the FPGA individually and using a more complicated interrupt mechanism to free the host threads and ensure asynchronous execution.
\vspace*{-3mm}
\section{Platform and Methodology}

ENEAC is developed and evaluated on the ZCU102 Development Platform and contains a heterogeneous scheduler to distribute workload between the on-board CPU and FPGA as well as custom hardware interrupt controllers and software interrupt mechanism to improve performance.

\subsection{The ZCU102 Development Platform}
\label{subsec:ZCU102_platform}
\begin{figure}[tp!]
\centering
\includegraphics[width=0.75\textwidth]{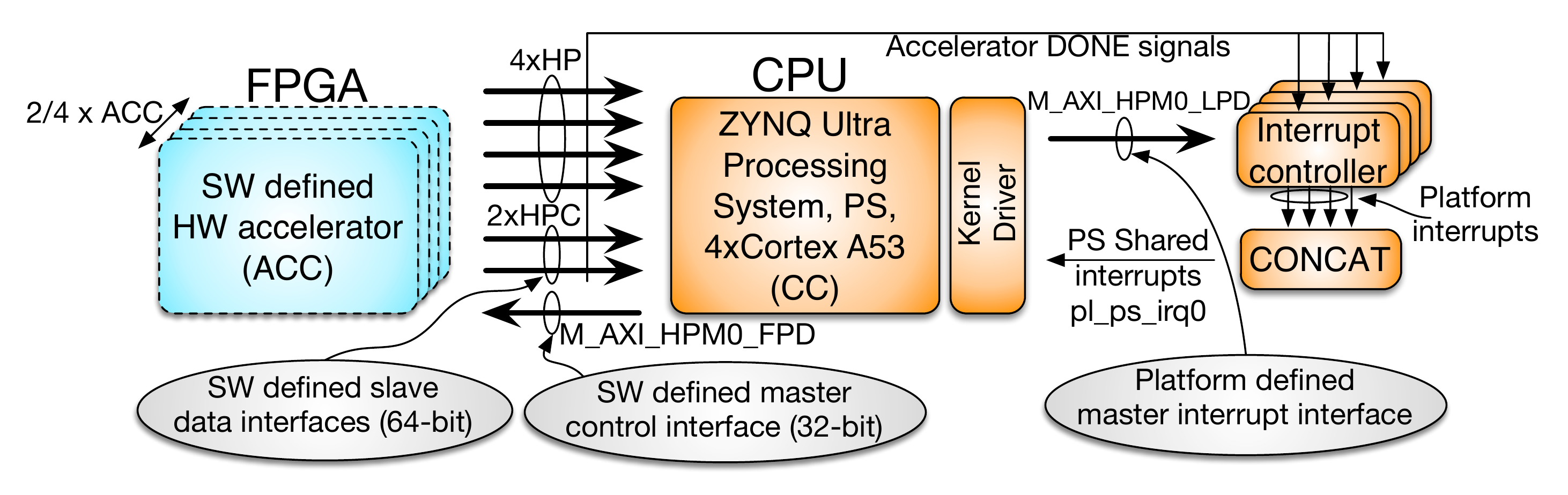}
\caption{SDSoC multiprocessing platform}
\vspace*{-4mm}
\label{fig:platform}
\end{figure}

The platform features the Xilinx Zynq Ultrascale+ SoC\cite{ZCU102}, which has a quad core ARM Cortex-A53, as well as an on-chip FPGA. Data transfer between the CPU and FPGA logic is done via AXI interface and the chip supports access via 4 High-Performance (HP) or 2 High-Performance-Cacheable (HPC) ports into CPU memory. Programming the FPGA is done via the SDSoC environment with optimised hardware accelerator implementations for the two benchmarks that are used in the evaluation. Both types of accelerator connections are evaluated using ENEAC with HPC ports being the preferred method of connection, since using the HP ports requires intermediate software data buffers from cacheable to non-cacheable memory.
\vspace*{-3mm}
\subsection{Custom interrupt generation}

A key component of ENEAC is the custom interrupt generation mechanism consisting of i) hardware interrupt generators, which connect to the CPU IRQ lines and indicate when each hardware accelerator is finished; and ii) software drivers, which catch the interrupts and wake the host thread (the thread in charge of offloading work to the FPGA). Figure~\ref{fig:platform} shows the hardware platform configurations, including the data access ports and the interrupt controllers. A key feature is that every FPGA accelerator has its own dedicated interrupt controller, interrupt driver and host thread so that each FPGA accelerator can perform independently. Moreover, the host thread does not waste CPU cycles waiting for the accelerator. 

\subsection{Software scheduler}
\begin{figure}[tp]
\centering
\includegraphics[width=0.75\textwidth]{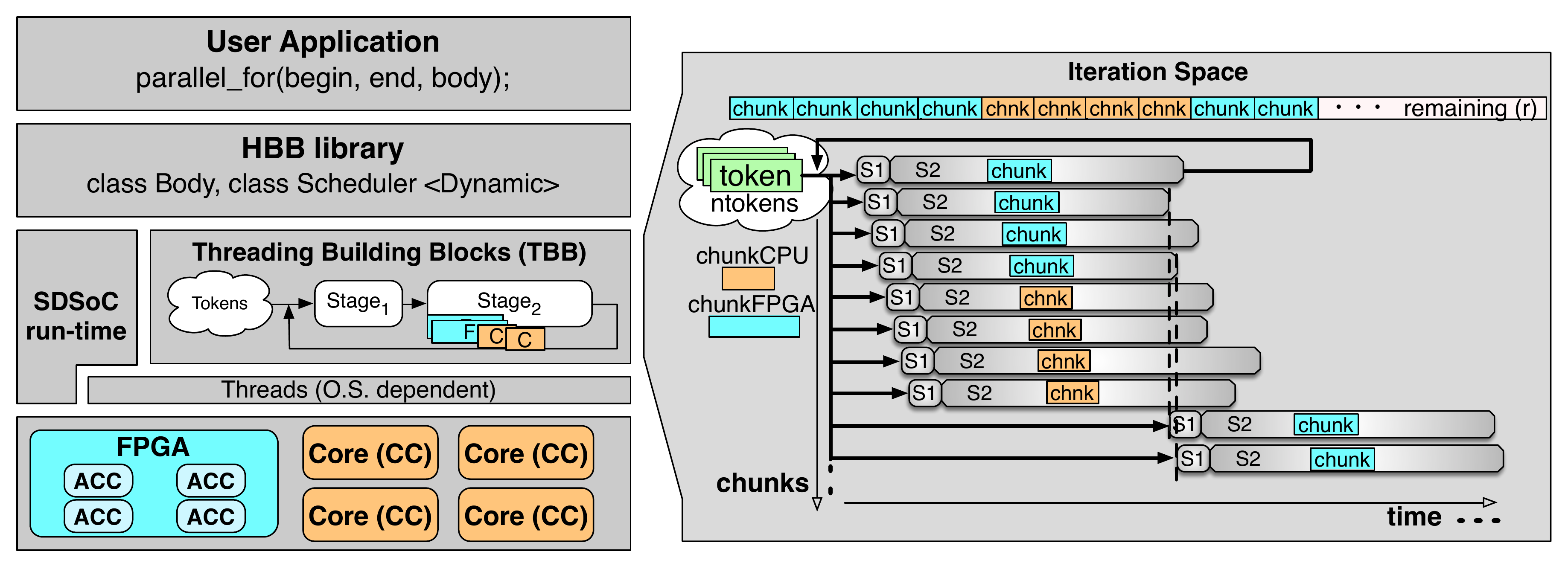}
\caption{The heterogeneous scheduling design}
\label{fig:scheduler}
\end{figure}

The custom heterogeneous software scheduler, that is part of ENEAC, builds on top of SDSoC and TBB libraries, and offers a \texttt{parallel\_for()} function template to run on heterogeneous CPU-FPGA systems. Fig.~\ref{fig:scheduler} shows the ZCU102 system with four FPGA accelerators (ACC) and four CPU cores (CC) as is used in the experimental evaluation. The left side shows the software stack that runs the workload, which includes the heterogeneous scheduler. It takes care of splitting the iteration space into chunks and processes each chunk on a CC or an ACC device. The right part illustrates how the internal engine managing the \texttt{parallel\_for()} works. The iteration space consists of the chunks already assigned to a processing unit and the remaining iterations waiting to be assigned. In the current implementation of the scheduler, called \textit{MultiDynamic}, the user specifies the ACC chunk size and the scheduler dynamically adapt the CC chunk size with the goal of maximizing the load balance. The scheduler supports the offload to each compute unit as soon as it becomes available, a feature particularly relevant for irregular workloads in which the execution time of a chunk of iterations can not be predicted at runtime.

\subsection{Implementation and workflow}
\label{subsec:implementation}
\begin{figure}[tp!]
\centering
\includegraphics[width=0.75\textwidth]{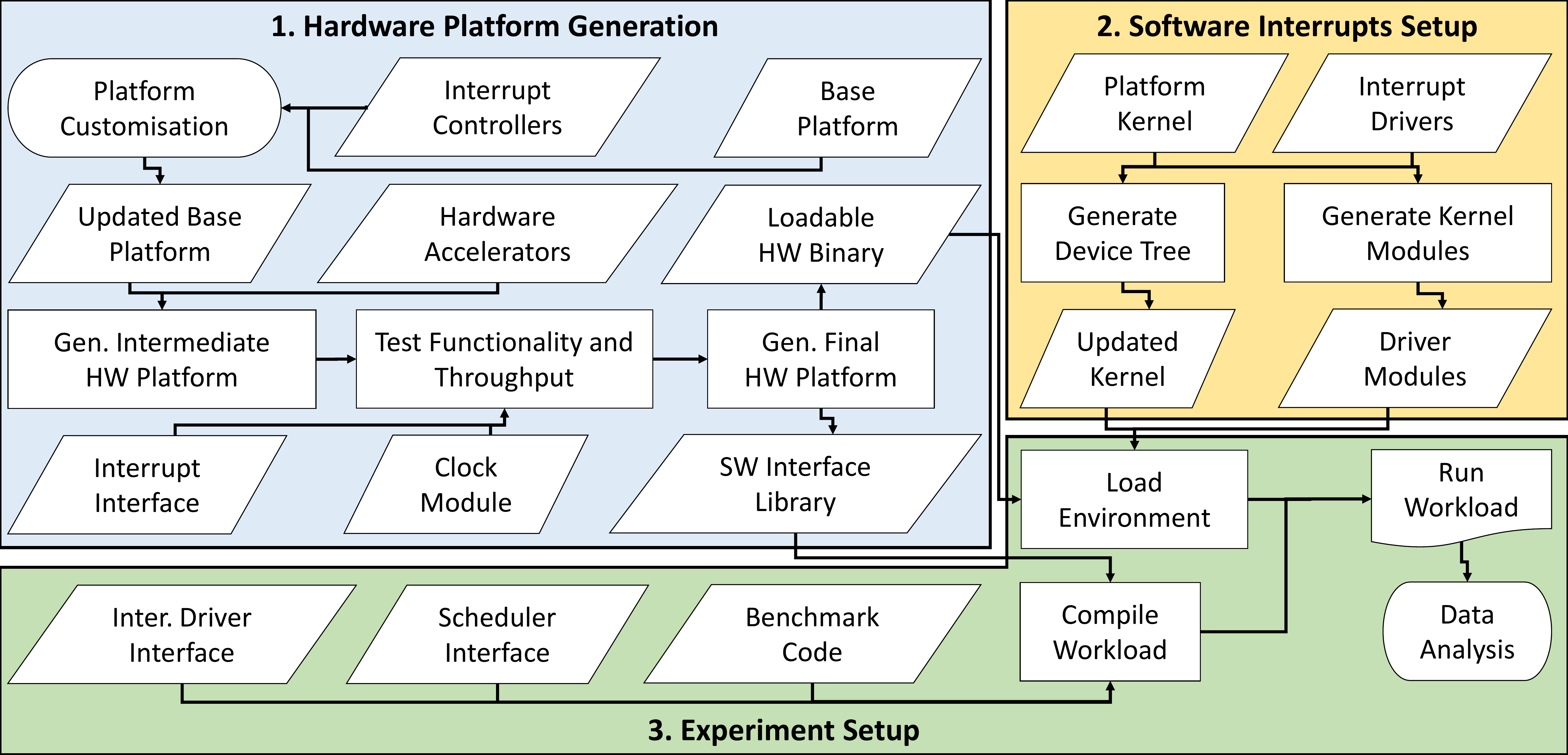}
\caption{ENEAC Workflow}
\vspace*{-4mm}
\label{fig:methodology}
\end{figure}

The three stages of the methodology are presented in Figure~\ref{fig:methodology}. Stage one consists of using the SDSoC tools to update the default platform with the interrupt controllers, introduce the application specific accelerator and ensure correct functionality. The final design uses a custom clock module, set to 200MHz for all configurations. The second stage involves updating the platform environment with the interrupt controller drivers. The experimental setup stage includes compiling the workload code with the scheduler software stack, the interrupt driver interface and the software interface library. Then the FPGA is loaded with the accelerator platform and the workload is executed, while collecting runtime data.
\vspace*{-3mm}
\section{Experimental evaluation}

ENEAC is evaluated with two distinct benchmarks - HOTSPOT and SPMM.

HOTSPOT is a stencil algorithm from the Rodinia benchmark collection, which estimates thermal dissipation on the surface of a chip by solving a series of differential equations. Workload size is altered by specifying the chip size. For the evaluation a chip size of 2048$\times$2048 points is used and the algorithm computes the rows in parallel
(2048 rows or iterations). 

SPMM multiplies a sparse matrix of 29957$\times$29957 with a randomised matrix of 29957$\times$100. The algorithm computes the matrix values per row so the iteration space contains 29957 iterations. This irregular benchmark aims to identify if ENEAC enables the accelerators and the CPU to work independently and reliably.

The performance measurement is standardised as throughput, given by compute objects per millisecond - temperatures for HOTSPOT and matrix rows for SPMM.  Seven distinct platform configurations are compared in the evaluation which are numbered with an ID from 1 to 7: (1) \textbf{4CC} uses just the four CPU cores to execute the workload; (2) \textbf{4HPACC} uses 4 FPGA accelerators connected through the HP ports to the PS (CPU); (3) \textbf{4HPCACC} uses HPC connected accelerators; (4)-(5) \textbf{4CC+4HPACC} distributes the workload between the 4 CPU cores and 4 HP connected accelerators without and with the hardware interrupts, respectively; and (6)-(7) \textbf{4CC+4HPCACC} refer to the last two configurations pairing the CPU with HPC connected accelerators, again without and with interrupts. For all configurations the performance is measured over a range of FPGA workload chunk sizes to identify the optimal workload distribution using the \textbf{MultiDynamic} scheduler presented earlier. 


\begin{figure*}[t!]
	\begin{minipage}[b]{1\textwidth}
      	\centering
      	\subfloat[HOTSPOT]{
      	\includegraphics[width=0.40\textwidth]{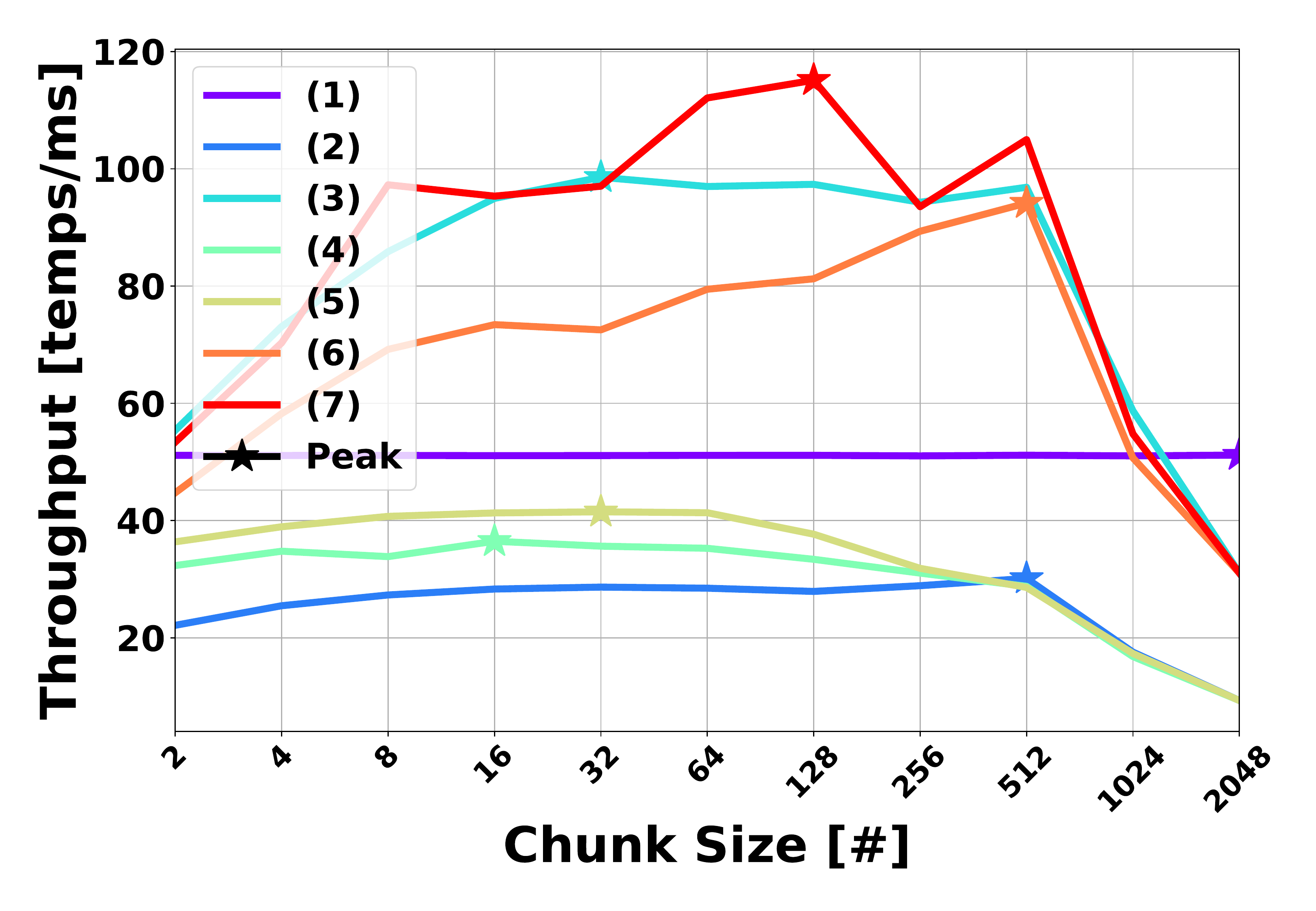}
      	\label{fig:HOTSPOT_performance}}
		\hfill
		\subfloat[SPMM]{
 		\includegraphics[width=0.40\textwidth]{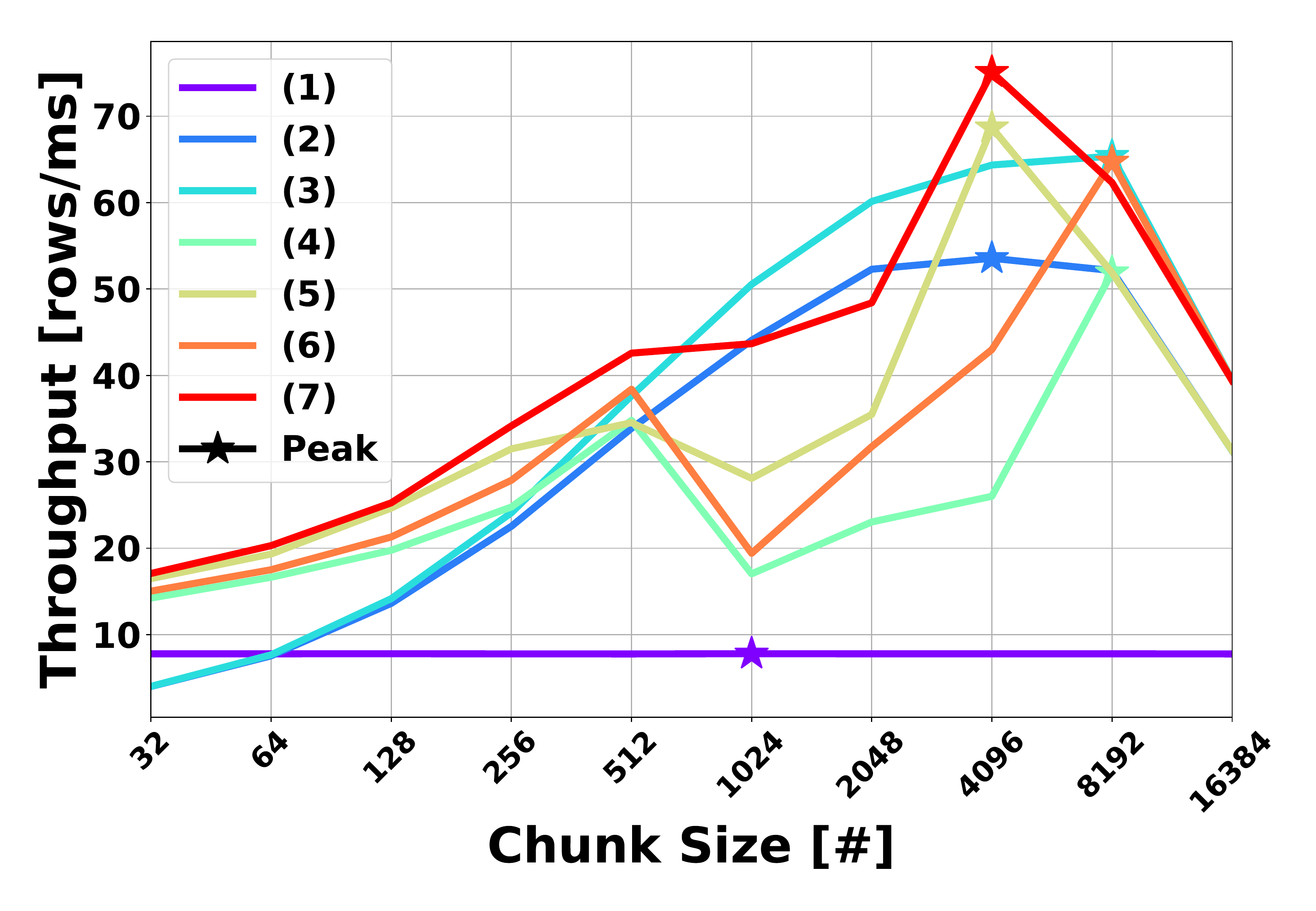}
      	\label{fig:SPMM_performance}}
      	\captionof{figure}{Benchmark performance on the ZCU102 Development Platform}
        \label{fig:all_performance}
    \end{minipage}
    
    \begin{minipage}[b]{1\textwidth}
        \centering
        \begin{tabular}{|c|l|c|c|c|c|}
\hline
\multirow{2}{*}{ID} & \multicolumn{1}{c|}{\multirow{2}{*}{Configuration}} & \multirow{2}{*}{Inter.} & \multirow{2}{*}{Sched.} & \multicolumn{2}{c|}{Peak Throughput} \\
                    & \multicolumn{1}{c|}{}                               &                         &                         & HOTSPOT            & SPMM            \\ \hline
(1)                 & 4CC                                                 & No                      & MD                      & 51.17              & 7.78            \\
(2)                 & 4HPACC                                              & No                      & MD                      & 30.13              & 53.55           \\
(3)                 & 4HPCACC                                             & No                      & MD                      & 98.55             & 65.38           \\
(4)                 & 4CC+4HPACC                                          & No                      & MD                      & 36.47              & 51.90           \\
(5)                 & 4CC+4HPACC                                          & Yes                     & MD                      & 41.49              & 68.67           \\
(6)                 & 4CC+4HPCACC                                         & No                      & MD                      & 94.15              & 64.71           \\
(7)                 & 4CC+4HPCACC                                         & Yes                     & MD                      & 115.11             & 75.09           \\ \hline
\end{tabular}
        \captionof{table}{Benchmark performance on the ZCU102 Development Platform}
        \label{tab:all_performance}
    \end{minipage}
\vspace*{-7mm}
\end{figure*}

Figure~\ref{fig:all_performance} and Table~\ref{tab:all_performance} show the results of the performance evaluation while using the \textbf{MultiDynamic} scheduler for heterogeneous computing. For both HOTSPOT and SPMM benchmarks, shown in \ref{fig:HOTSPOT_performance} and \ref{fig:SPMM_performance} respectively, it can be observed that the highest throughput point is achieved using the final configuration using all 4 CPU Cores and 4 HPC-connected FPGA accelerators. Comparison between configurations (6) and (7) and configurations (4) and (5) reveals that in both benchmarks using the custom interrupt mechanism improves resource utilisation and increases the platform throughput regardless of the accelerator data ports. For both benchmarks the optimal point for the workload distribution varies between configurations and a sharp decrease in throughput can be observed when more that 1/4 of the workload is scheduled per accelerator - 512 chunk size for HOTSPOT and 8192 chunk size for SPMM.

The results obtained using ENEAC show that utilising the heterogeneous scheduler to distribute the workload between the CPU and FPGA with the custom interrupt mechanism on the Xilinx ZCU102 development board produces the highest throughput across the hardware configurations for both benchmarks included in the evaluation. Using the workflow results in 124.96\%, 16.80\% and  22.26\% increase in throughput for the HOTSPOT benchmark when compared to just using 4CPU cores, only 4HPC FPGA accelerators, and 4CPU + 4HPC FPGA  accelerators without the interrupt mechanism, respectively. The irregular SPMM benchmark also shows a significant improvement of 865.17\%, 14.85\%  and 16.04\% over the three equivalent hardware configurations.

\vspace*{-3mm}
\section{Conclusion}

This paper presents ENEAC, a custom methodology to optimally distribute workloads on a complex heterogeneous computing platform between the multicore CPU and the on-board FPGA. Multiple FPGA hardware configurations are explored and ENEAC manages to successfully integrate custom hardware accelerators and interrupt mechanism to improve workload execution, compared to just using the hardware accelerators without help from the CPU, by 16.80\% and 14.85\%  for the HOTSPOT and SPMM benchmarks respectively. \ignore{As a result the methodology also helps decrease energy usage, particularly for SPMM by 13.88\%.} 

Future work involves optimising and evaluating a more advanced custom scheduler on the platform, which identifies the optimal workload distribution automatically, without the need to manually set the chunk sizes and also focusing specifically on optimising energy efficiency in addition to throughput. A larger set of benchmarks will be used to demonstrate the general applicability of the methodology. ENEAC is open-source and can be accessed at~\cite{ENEAC}.

%
%
%
\bibliographystyle{splncs04}
\bibliography{references}

\begin{thebibliography}{1}
\providecommand{\url}[1]{\texttt{#1}}
\providecommand{\urlprefix}{URL }
\providecommand{\doi}[1]{https://doi.org/#1}

\bibitem{ENEAC}
{ENergy Efficient Adaptive Computing with multi-grain heterogeneous
  architectures (ENEAC)}. \url{https://github.com/eejlny/ENEAC}, accessed:
  2018-10-15

\bibitem{HSA}
{HSA Foundation}. \url{www.hsafoundation.com}, accessed: 2018-03-05

\bibitem{ZCU102}
{Xilinx Zynq UltraScale+ MPSoC ZCU102 Evaluation Kit}.
  \url{https://www.xilinx.com/products/boards-and-kits/ek-u1-zcu102-g.html},
  accessed: 2018-10-15

\bibitem{micro10}
Chung, E.S., Milder, P.A., Hoe, J.C., Mai, K.: Single-chip heterogeneous
  computing: Does the future include custom logic, {FPGAs}, and {GPGPUs}? In:
  Micro '10 (Dec 2010)

\bibitem{esmaeilzadeh2011dark}
Esmaeilzadeh, H., Blem, E., Amant, R.S., Sankaralingam, K., Burger, D.: Dark
  silicon and the end of multicore scaling. In: 2011 38th Annual international
  symposium on computer architecture (ISCA). pp. 365--376. IEEE (2011)

\bibitem{nunez2019simultaneous}
Nunez-Yanez, J., Amiri, S., Hosseinabady, M., Rodr{\'\i}guez, A., Asenjo, R.,
  Navarro, A., Suarez, D., Gran, R.: Simultaneous multiprocessing in a
  software-defined heterogeneous fpga. The Journal of Supercomputing
  \textbf{75}(8),  4078--4095 (2019)

\bibitem{shafique2014dark}
Shafique, M., Garg, S., Mitra, T., Parameswaran, S., Henkel, J.: Dark silicon
  as a challenge for hardware/software co-design: Invited special session
  paper. In: Proceedings of the 2014 International Conference on
  Hardware/Software Codesign and System Synthesis. p.~13. ACM (2014)

\bibitem{fpga10}
Tsoi, K.H., Luk, W.: Axel: A heterogeneous cluster with {FPGAs} and {GPUs}. pp.
  115--124. FPGA '10 (2010)

\end{thebibliography}
\end{document}